# Environmental Impact of Bundling Transport Deliveries Using SUMO

## Analysis of a cooperative approach in Austria

Aso Validi, Nicole Polasek, Leonie Alabi, Michael Leitner, Cristina Olaverri-Monreal*

Chair Sustainable Transport Logistics 4.0
Johannes Kepler University
Linz, Austria
aso.validi@jku.at, cristina.olaverri-monreal@jku.at

*Abstract*—**Urban Traffic is recognized as one of the major CO$_2$ contributors that puts a high burden on the environment. Different attempts have been made for reducing the impacts ranging from traffic management actions to shared-vehicle concepts to simply reducing the number of vehicles on the streets. By relying on cooperative approaches between different logistics companies, such as sharing and pooling resources for bundling deliveries in the same zone, an increased environmental benefit can be attained. To quantify this benefit we compare the CO$_2$ emissions, fuel consumption and total delivery time resulting from deliveries performed by one cargo truck with two trailers versus by two single-trailer cargo trucks under real conditions in a simulation scenario in the city of Linz in Austria. Results showed a fuel consumption and CO2 emissions reduction of 28% and 34% respectively in the scenario in which resources were bundled in one single truck.**

*Keywords-component; transport logistics, CO2 emissions, fuel consumption, travel time*

## I. INTRODUCTION

The transportation sector is the third largest consumer of energy and the largest consumer of petroleum products and as a consequence, one of the greatest contributors to air pollution worldwide [1].

Considering the increasing demand for transport and the growth of e-commerce it is crucial to analyze traffic and provide solutions to reduce its environmental impact. Measures related to vehicular emissions provide insights into the scope of the problem. The acquisition of these data can be costly, risky and require a lot of time and effort [2]. Therefore, a common approach is to use controllable test environments that include relevant information pertaining to the use case to be studied. Traffic simulation provides a framework for the representation of complex, realistic traffic situations that can be useful in evaluating a specific traffic situation or testing new technological applications [3].

In this paper we define, simulate and analyze two scenarios for transport delivery in the city of Linz in Austria. To this end we rely on the Simulation of Urban Mobility (SUMO) framework [4].

We compare the CO2 emissions, fuel consumption and time on the road resulting from deliveries performed by a cargo truck with two trailers and by two cargo trucks with one trailer on a defined route with delivery stops in the city of Linz, Austria.

The remaining parts of this paper are organized as follows: Section II introduces the related work reported in the literature. Section III presents the developed methodology. The results from the analysis of the defined scenarios are presented in Section IV. In Section V, conclusions and prospective work in line with our research are described.

## II. RELATED WORK

In recent years, the amount of research work conducted that suggests cooperation and consolidation for improving transportation operations efficiency has been increasing [5]. Research on truckload utilization, fuel consumption, costs, timing and time management as well as Green House Gas (GHG) and CO$_2$ emission mitigation has reached very similar conclusions while adopting different methodologies [6]. In this section, some related published works are reviewed and summarized and the gap in literature to which our research is trying to contribute is highlighted.

The authors in [7] have developed an optimization model based on Discrete Time-Based Shipment Consolidation (DTB-SCL) to quantify the benefits of sharing shipments in terms of both cost and CO$_2$ emission. This research work concludes and suggests that consolidating deliveries will decrease the environmental impact of the transportation network through a mathematical optimization model.

A model is developed to enhance vehicle fleet utilization in [8]. This optimization model takes into account different operational factors such as cargo size, loading sequence, truck capacity, arrival time and carbon emission in utilizing the fleet of vehicles. A MATLAB-based methodology is adopted for solving this model and simulating the developed systems. The final results suggest that optimizing truckload has a positive effect on CO$_2$ emission and fuel consumption.

A EURO Working Group on Transportation-funded project [9] introduces a case study focused on the implications of urban freight consolidation centers and their positive effects on urban logistics and deliveries in city areas. The developed model analyzes the effects of consolidation on reducing city traffic, fuel consumption and $CO_2$ emissions. The results suggest that consolidation deliveries significantly decrease $CO_2$ emissions, fuel consumption and traffic of delivery trucks in urban areas.

The concept of "combinatorial auctions" is the main focus of [10]. This delivery mechanism is built based on submitting bids for transportation lane contracts when individual or combined bundles are available. The results of this project suggest that consolidated bids increase truckload utilization and mitigate the negative effects of freight transportation, emission and traffic being amongst them.

Although [11] also discusses and proposes pooling, the main point of discussion is reducing $CO_2$ emission from logistics activities within the supply chain. The developed mathematical optimization methodology is a logistics network model that examines pooling throughout the network by considering different scenarios. Results suggest that $CO_2$ emission and the frequency of deliveries can both be reduced, affecting thus fuel consumption.

A critical review of related literature reveals that a considerable number of studies and research projects have focused on truckload pooling and consolidation in transportation to initially minimize the costs and reduce the environmental impact of transportation in both freight and city logistics contexts. Mathematical optimization is the dominant method adopted. To the best of our knowledge, SUMO has never been adopted for this purpose. Use of simulation via SUMO has capabilities beyond an optimization model that has not been explored before. Furthermore, single and double trailer truckloads have not been reported to be studied and compared in a city logistics delivery context. Hence, this paper contributes to the body of knowledge by exploring both of these gaps in literature.

## III. METHOD

As previously mentioned, we adopted a simulation-based analysis to investigate the potential environmental benefits of bundling transport deliveries in one truck instead of two. To this end we used the two-dimensional microscopic traffic simulation. SUMO, which is a time-discrete traffic simulation platform that allows modelling of noise and emissions, is open-source and has a proven track record with other projects [3, 12,13].

In addition to SUMO we adopted the Traffic Control Interface (TraCI) [14]. TraCI uses a Transmission Control Protocol (TCP)-based client/server architecture to provide access to SUMO. Each vehicle in the SUMO network is defined by a unique identifier, departure time, and route.

Within this study, two scenarios were defined:

- *Scenario I* corresponded to a single truck consisting of two containers (trailer) and supplying two delivery points.

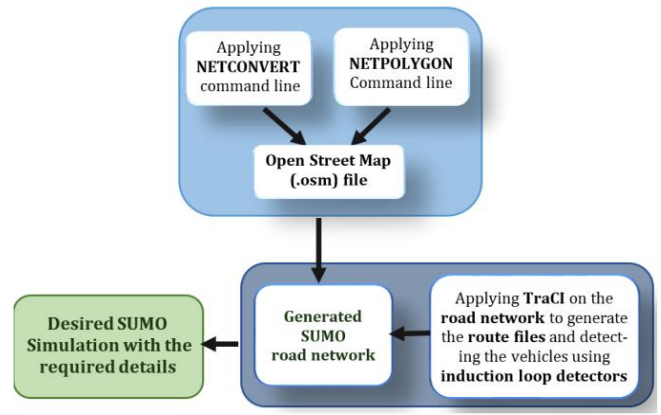

Figure 1. Sumo simulation process to generate traffic and road network [21]

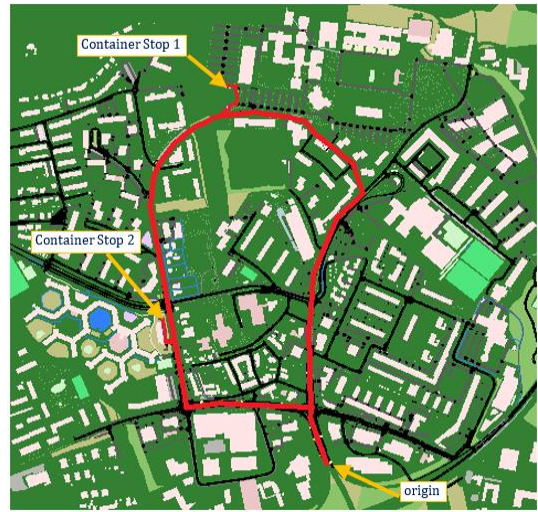

Figure 2. Illustration of the imported network showing the container stops (delivery points), origin (starting point) and the defined route (1.95 km)

- *Scenario II* corresponded to two trucks, each equipped with one container, each truck having one delivery point.

In order to generate the simulation network, a part of the city of Linz that included the area around the Johannes Kepler University, Altenberger Straße, and a section of the motorway were imported into SUMO from OpenStreetMap (OSM) [15]. NETCONVERT [16] and POLYCONVERT [17] were then applied to the OSM data to generate the road network in SUMO. In order to generate the route files and define the vehicle types, the *vType* element was applied on the network.

In addition to the road and traffic network files, files for analyzing the emissions were generated to simulate delivery points (container stops) [18] and to write values to the output as soon as the vehicle was detected (induction loop detectors [19]). In these additionally generated files, all the simulated trucks were implemented with the SUMO emission class average

---

**Algorithm 1:** Applying TraCI for generating the route and induction loop detectors files

---

**input:** Road Networks r;
**output:** Route file ru; Detection file; add
**define**: Number of vehicles i; Number of time steps n

1  **function** *generating the related route file and vehicle types ():*
2      random. seed (s)
3      N ← n
4      writing on a .xml file as routes:
5          **output** route file with different vType, route edges
                      And vehicles id
6      vehNr ← 0
7      **for** *i in range(N):*
8          **if** *random.uniform (0, 1) < single truck*:
9              **end for**
10             **output** '*<generating the vehicle types with
                          different attributes/>' format (i,
                          vehNr),* file=routes
11             vehNr += 1
12         **end if**
13         **if** *random.uniform(0, 1) < double truck:*
14             **end for**
15             **output** '*<generating the vehicle types with
                          different attributes/>' format (i,
                          vehNr),* file=routes
16             vehNr += 1
17         **end if**
18     **end for**
19     **output** route file
20 **end function**

21 **function** *running the simulation and detecting vehicles ():*
22     **while** *traci.simulation.getMinExpectedNumber ( ) > 0:*
23         traci.simulationStep ()
24     **end while**
25     **for** *i in range (N):*
26         **output** traci.simulation.getTime ()
27     **end for**
28     **for** *det in traci.inductionloop.getIDList ():*
29         **output** file with details for all induction detector
                      loops
30     **end for**
31 **end function**

---

gasoline-driven heavy duty vehicle "HBEFA3/HDV_G", adhering to the vehicular pollutant emissions based on the database application HBEFA [20]. The main steps for generating the road and traffic network in SUMO simulation are presented in Fig. 1 [21]. In order to generate the route files from the road network and sense the vehicles passing by the induction loop detectors we adopted TraCI. The procedure with the instruction of adopting TraCI is depicted in Algorithm 1.

The network simulation, the defined route and the two container stop locations are presented in Fig. 2. Within the simulation two container stops were defined and generated as the delivery points. Three trucks, one with two containers (Scenario I) and two single trucks (Scenario II) were defined within the exit number 15 of the motorway A7 as their starting point (presented as the origin in Fig. 2). The delivery points are the Spar (University) and the Spar (Dornach) (grocery stores). The truck with two containers supplied both Spar supermarkets,

Defined induction loop detectors were implemented in frequencies of 50s to acquire data related to emission and fuel consumption. The speed limits of the trucks were 50km/h in city limits and 80 km/h on the motorway and country roads. The related speed map is presented in Fig. 3, which shows the maximum speed limit on different routes in the network.

The simulation parameters for both routes and the additional files are presented in Table 1. The network was edited in terms of size and unused roads, such as pedestrian routes and activating the traffic lights by using NETEDIT [22]. The position of the traffic lights extracted from the network is illustrated in Fig. 4.

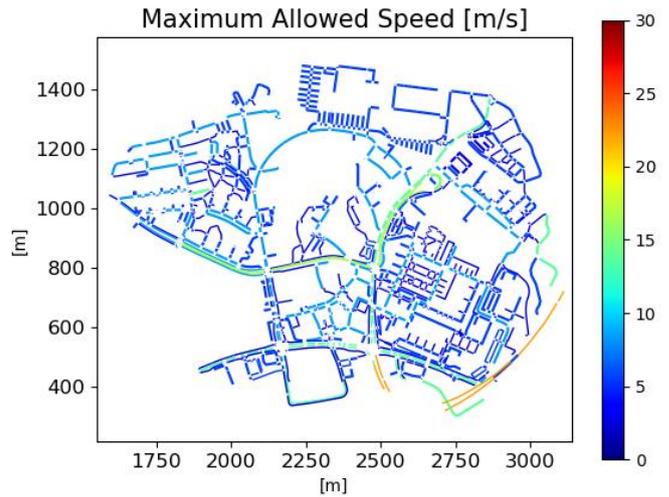

Figure 3.   Speed map range of the simulated network

TABLE I.        DEFINED SUMO SIMULATION PARAMETERS

| SUMO Parameters | | |
|---|---|---|
| **Route file** | ***vClass*** | Scenario I → a truck with two containers |
| | | Scenario II → two single trucks each with one container |
| | ***maxSpeed*** | 15 m/s |
| | ***minSpeed*** | 5 m/s |
| **Additional file** | ***Freq a[1]*** | 50 s |
| | ***containerStop*** | Scenario I → 2 stops |
| | | Scenario II → 1 stop |
| | ***emissionClass*** | HBEFA3/HDV_G |

[1] attribute of Induction Loop Detectors

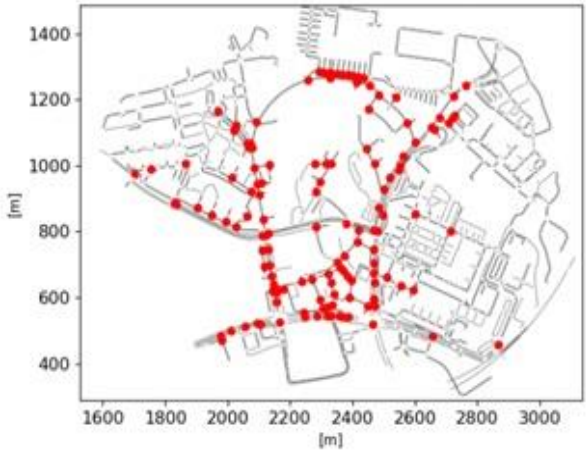

Figure 4.   Location of the traffic lights in the selected route

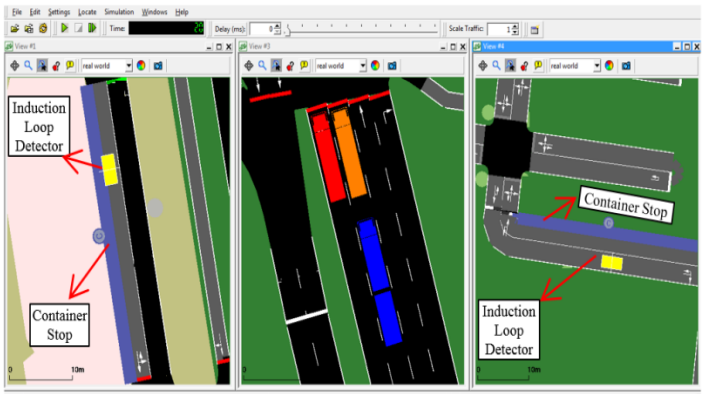

Figure 5.   The generated simulation with the defined trucks: single trucks (orange & red), double truck (blue), container stops and induction loop detectors.

TABLE II.   RESULTING EMISSION AND FUEL CONSUMPTION FROM THE PERFORMED ANALYSIS

| | $CO_2$ Emission (cumulated) | | Fuel Consumption (cumulated) | |
|---|---|---|---|---|
| **Scenario I** (a truck with two container) | 2725 mg/s 2.7 kg/s | | 1262ml/s | |
| **Scenario II** (single trucks each with one container) | 2554 mg/s 2.6 kg/s | sum = 5192mg/s sum = 5.5 kg/s | 1171 ml/s | Sum= 2269 ml/s |
| | 2938 m/s 2.9 kg/s | | 1098 ml/s | |

Fig. 5 illustrates the generated two single-container trucks (red and orange), the truck with two containers (blue), the implemented detectors and the container stops. In order to simulate the delivery points, the route priorities were adjusted so that the supermarkets could also be accessed by side streets rather than directly from the front door. In this case we changed the priority of the side streets and assigned them the same priority level '10' of a main street. We also adjusted the settings as necessary to define which vehicles were allowed to drive on the roads and which were not.

## IV. ANALYSIS RESULTS

In this section, we show the comparative analysis of the simulation results for the two defined scenarios.

The emissions produced by the truck with two containers delivering to two locations (Scenario I) totaled 2,7 kilograms of $CO_2$ , much less compared to the 5,5 kilograms from the two trucks in Scenario II (Table 2), the emissions reduction resulting from bundling goods into one vehicle being  34%.

As depicted in Fig. 3, the speed limit of the routes for the two separate delivery trucks were 60 km/h, higher than the speed limit for the double-container/trailer truck (50 km/h), this fact contributing to the resulting higher emissions of Scenario II.

Fig. 6 depicts the stops and waiting times for Scenario I and Scenario II. During the time period from 360 to 390 seconds the horizontal bar in the truck with two containers shows that it stopped at a traffic light for around 30 seconds. The graphic shows the highest emissions at the beginning and at the end of the routes, because at these times the trucks are driving on the motorway.

Fig. 7 shows the comparison between the fuel consumption (ml/s) for the two single container trucks (Scenario II) and the truck with two containers (Scenario I). A fuel savings of 28% by using one single truck is depicted.

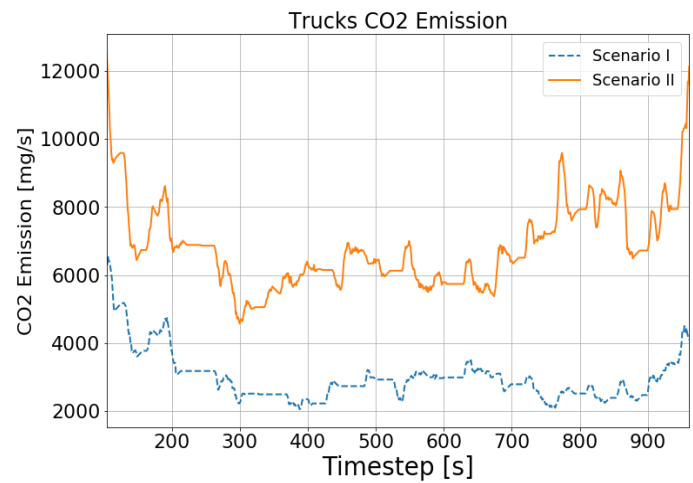

Figure 6.   $CO_2$ emissions resulting from double container truck (Scenario I) and single container trucks (Scenario II)

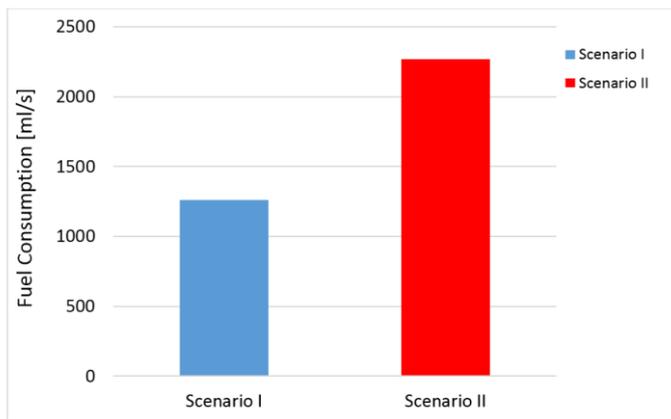

Figure 7. Comparison between fuel consumption of the double container truck Scenario I) and the two single container trucks (Scenario II)

## V. DISCUSSION AND FUTURE WORK

To alleviate traffic in our roads it is essential that companies cooperate to use fewer vehicles. As shown in this work, the implementation of cooperative approaches to bundle cargo would benefit the environment by reducing a considerable amount of $CO_2$ emission and fuel consumption. A timely delivery resulting from additional stops could be ensured through accurate time planning and management. Until now, getting deliveries faster has been prioritized. However, ignoring the consequences of transport regarding pollution is not an option and every possible solution should be investigated. Therefore, the implementation of shared concepts, where store deliveries are consolidated and delivered by fewer vehicles would be a very good step forward in reducing emissions. Future work will aim at extending these scenarios and approaches by evaluating the emission classes in SUMO and developing further emission models.

## ACKNOWLEDGMENT

This work was supported by the FFG project "Zero Emission Roll-Out - Cold Chain Distribution_877493" and the Austrian Ministry for Climate Action, Environment, Energy, Mobility, Innovation and Technology (BMK) Endowed Professorship for Sustainable Transport Logistics 4.0.

## REFERENCES

[1] McKinnon A. Decarbonizing logistics: Distributing goods in a low carbon world. Kogan Page Publishers; 2018 Jun 3.Kouvelis, P., Dong, L., Boyabatli, O., & Li, R. (2011). Handbook of integrated risk management in global supply chains (Vol. 1). John Wiley & Sons.

[2] Kouvelis P, Dong L, Boyabatli O, Li R. Handbook of integrated risk management in global supply chains. John Wiley & Sons; 2011 Oct 26.

[3] Biurrun-Quel C, Serrano-Arriezu L, Olaverri-Monreal C. Microscopic driver-centric simulator: Linking Unity3d and SUMO. InWorld Conference on Information Systems and Technologies 2017 Apr 11 (pp. 851-860). Springer, Cham.

[4] Pattberg, B. Institute of transportation systems - SUMO – simulation of urban mObility. Available at: http://www.dlr.de/ts/en/desktopdefault.aspx/tabid-9883/16931_read41000/ [accessed: 28 January 2020]

[5] Sanchez M, Pradenas L, Deschamps JC, Parada V. Reducing the carbon footprint in a vehicle routing problem by pooling resources from different companies. NETNOMICS: Economic Research and Electronic Networking. 2016 Jul 1;17(1):29-45.

[6] Yao X, Cheng Y, Song M. Assessment of collaboration in city logistics: From the aspects of profit and CO2 emissions. International Journal of Logistics Research and Applications. 2019 Nov 2;22(6):576-91.

[7] Ülkü MA. Dare to care: Shipment consolidation reduces not only costs, but also environmental damage. International Journal of Production Economics. 2012 Oct 1;139(2):438-46.

[8] Wong EY, Tai AH, Zhou E. Optimising truckload operations in third-party logistics: a carbon footprint perspective in volatile supply chain. Transportation Research Part D: Transport and Environment. 2018 Aug 1;63:649-61.

[9] Daniela P, Paolo F, Gianfranco F, Graham P, Miriam R. Reduced urban traffic and emissions within urban consolidation centre schemes: The case of Bristol. Transportation Research Procedia. 2014 Jan 1;3:508-17.

[10] Mesa-Arango R, Ukkusuri SV. Benefits of in-vehicle consolidation in less than truckload freight transportation operations. Procedia-Social and Behavioral Sciences. 2013 Jun 7;80:576-90.

[11] Ballot E, Fontane F. Reducing transportation CO2 emissions through pooling of supply networks: perspectives from a case study in French retail chains. Production Planning & Control. 2010 Sep 1;21(6):640-50.

[12] Krajzewicz D, Erdmann J, Behrisch M, Bieker L. Recent development and applications of SUMO-Simulation of Urban MObility. International Journal On Advances in Systems and Measurements. 2012 Dec 3;5(3&4).

[13] Olaverri-Monreal C, Errea-Moreno J, Díaz-Álvarez A, Biurrun-Quel C, Serrano-Arriezu L, Kuba M. Connection of the SUMO microscopic traffic simulator and the Unity 3D game engine to evaluate V2X communication-based systems. Sensors. 2018 Dec;18(12):4399.

[14] TraCI (November 2019) "Multiple clients" available: https://sumo.dlr.de/docs/TraCI.html [accessed 06 February 2020].

[15] OpenStreetMap (2019). "Open Street Map", available: https://www.openstreetmap.org/about abgerufen. [accessed 15December 2019].

[16] NETCONVERT ( December 2019) "Further supported Data Formats", available: https://sumo.dlr.de/docs/NETCONVERT.html [accessed 07 February 2020].

[17] POLYCONVERT (December 2019) "Configuration", available: https://sumo.dlr.de/docs/POLYCONVERT.html [accessed 07 February 2020].

[18] Container Stop (December 2019) "Specification/Containers", available: https://sumo.dlr.de/docs/Specification/Containers.html [accessed 07 February 2020].

[19] Induction Loop Detectors (September 2019) "Instantiating within the Simulation", available: https://sumo.dlr.de/docs/Simulation/Output/Induction_Loops_Detectors_(E1).html [accessed 07 February 2020].

[20] Emission Classes-SUMO (September 2019) "Models/Emissions", available: https://sumo.dlr.de/docs/Models/Emissions.html [accessed 07 February 2020].

[21] Validi A, Ludwig T, Hussein A, Olaverri-Monreal C. Examining the impact on road safety of different penetration rates of vehicle-to-vehicle communication and adaptive cruise control. IEEE Intelligent Transportation Systems Magazine. 2018 Sep 12;10(4):24-34.

[22] NETEDIT (November 2019) "Edit Modes", available: https://sumo.dlr.de/docs/NETEDIT.html [accessed 06 February 2020].